**Genomic Encryption of Biometric Information for Privacy-Preserving Forensics**

Taeho Jung[1], Ryan Karl[1], Geoffrey H. Siwo[2,3,4*]

[1]Department of Computer Science & Engineering, University of Notre Dame, IN, USA
[2]Center for Research Computing, University of Notre Dame, IN, USA
[3]Eck Institute for Global Health, University of Notre Dame, IN, USA
[4]Department of Biological Sciences, University of Notre Dame, IN, USA

Corresponding author: gsiwo@nd.edu

**Abstract**
DNA fingerprinting is a cornerstone for human identification in forensics, where the sequence of highly polymorphic short tandem repeats (STRs) from an individual is compared against a DNA database. This presents significant privacy risks to individuals with DNA profiles in the database due to hacking by malicious attackers who may access the data and misuse it for secondary purposes. In this paper, we propose a novel cryptographic framework for jointly encrypting DNA-based fingerprints (STRs) with other biometric data, for example, facial images, such that the STRs and biometrics information of an individual are revealed only when a positive match is found, i.e. the STRs act as decryption keys. Specifically, when a search is performed on the encrypted database using STR sequences of an individual in the database, a perfect match generates the facial image and/ or other biometrics of the individual while the lack of a match returns a null result. By jointly encrypting DNA fingerprints and other biometrics using the unique STRs generated keys, our approach ensures perfect privacy of the encrypted information with decryption of only the record with STRs matching the query. This safeguards the information of other individuals in the same database. The proposed approach can also be used to securely authenticate the identity of individuals or biological material in scenarios beyond forensics including tracking the identity of samples for clinical genetics and cell therapies.

**Introduction**
Forensic DNA databases containing highly polymorphic short tandem repeats (STR) in the human genome have played a critical role in uniquely identifying suspects in crimes as well as identifying missing persons and the remains of individuals in disasters [1,2]. Forensic investigators in the United States and other countries have previously relied on STR-based national DNA databases such as the Combined DNA Index System (CODIS) maintained by the Federal Bureau of Investigations (FBI) [3]. Such national DNA databases have been mainly limited to individuals who have been arrested, charged or convicted. Typically, when investigating a crime scene, law enforcement officers often try to identify suspects by matching the DNA acquired from the site against the forensic DNA databases. There are privacy concerns in this process as the DNA database needs to be in an unencrypted form during the search process. Due to the insider threats (e.g., employees misusing their privileges, employees' devices being compromised by attackers), such a system has the risk of leakages of raw DNA fingerprints and the identity information associated with them, and this may lead to serious data breaches with catastrophic consequences. Furthermore, the biased racial composition of forensic databases such as CODIS could exacerbate racial profiling and, breaches or misuse of the data could disproportionately impact African Americans [4].

Besides forensic databases, the exponential growth in consumer genetics has led to an explosion of DNA sequences that can be used for human identification with over 26 million consumers with their DNA profiled by such companies [5]. In the search for suspects in high-profile criminal investigations, authorities have turned to these companies for access to a more extensive database of DNA sequences [6]. This situation creates a paradox for the companies who are caught between cooperating with law enforcement vs. their customers who never consented for such access and are concerned that this creates a backdoor that could be misused. Well-meaning consumers are also torn-between consenting for the use of their DNA information in criminal investigations vs. risks that their data may be used to target them at individual or group level. While in principle solving criminal cases is highly desirable, the ability of third-parties such as law enforcement to broadly access unique DNA fingerprints, especially those not relevant to a criminal investigation, infringes on individual rights to privacy and creates loopholes for adversaries including hackers. As millions of genomes continue to be sequenced, this also poses an increasing threat to national security as foreign actors could use similar loopholes to access genetic information of millions of citizens.

The current means of DNA matching does not allow for privacy-preserving querying of the forensic DNA databases which threatens the privacy of millions of individuals. It also discourages cooperation of consumer genetic companies in crime investigations, which is beneficial to society in solving crimes or identifying missing persons. To address this problem, we propose a novel cryptography-based framework to enable privacy-preserving queries of DNA databases that remain encrypted at all times, including to authorities interacting with it. Furthermore, the framework jointly encrypts DNA fingerprints with facial appearance (or other biometrics). A positive match leads to the generation of a facial image (or biometrics) of the suspect that can be directly validated by a non-expert to confirm the identity of the suspect. The ability of the system to generate biometric information, for example, faces can also be used to corroborate the presence of a suspect at a scene through witnesses or video evidence. It could also enable a suspect to verify that their DNA has produced a positive match without putting all trust in the authorities performing the search.

**Leveraging the entropy of STRs to encrypt biometric data**
We propose a radically novel approach to address the privacy risks discussed above. In a nutshell (Fig. 1), we use STR genetic loci that contain sufficient entropy for generating cryptographic keys. We then use these STR-derived cryptographic keys to encrypt the images and any associated identity information or biometrics. The STRs are known to have enough entropy that can be leveraged to generate cryptographic keys of at least 80 bits which provide sufficient security guarantees [7]. Specifically, we leverage the current National Institute of Standards (NIST) defined STRs for DNA profiling which are based on 20 core highly-polymorphic STR loci, a gold-standard for uniquely identifying individuals [3]. The information density of these loci given the diversity in the human population is so high that they cannot be computed in reasonable time by brute force [7]. Thus, it is possible to encrypt the face/identity data with 80-bit security level using standard symmetric-key encryption such as the Advanced Encryption Standard (AES) algorithm. The first step in implementing our framework is constructing an encrypted DNA/ biometric database in which the STRs and other biometric data of each individual is encrypted using keys generated from their STR sequence. Typical forensic databases such CODIS do not contain other biometric data but by leveraging our proposed joint encryption of STRs with biometrics, the privacy of biometrics (if available) is guaranteed. In addition, the proposed framework can also still be used where other biometrics are lacking.

To use the system in the real world, law enforcement officers collect the DNA samples from the crime scene and sequence the STRs of the unknown suspects, from which the keys for encryption can be generated using the same algorithms that were applied to encrypt the database. The keys from the suspect can then be used to search the encrypted DNA database for the decryptable ciphertext (i.e., the ciphertext that was generated with the key), which will contain the face and identity information of the person with the same DNA/STR sampled from the crime scene. For easily learning whether the decryption is successful, a predefined string can be added (e.g., a certain number of zero bits) as a preamble to all face/identity records. The two can then be encrypted together in case the decryption results into incorrect results without errors by coincidence (e.g., when the final block is properly padded by coincidence in AES algorithms). Fig. 2 demonstrates a toy example of a match and a non-match as a simple demonstration of the framework.

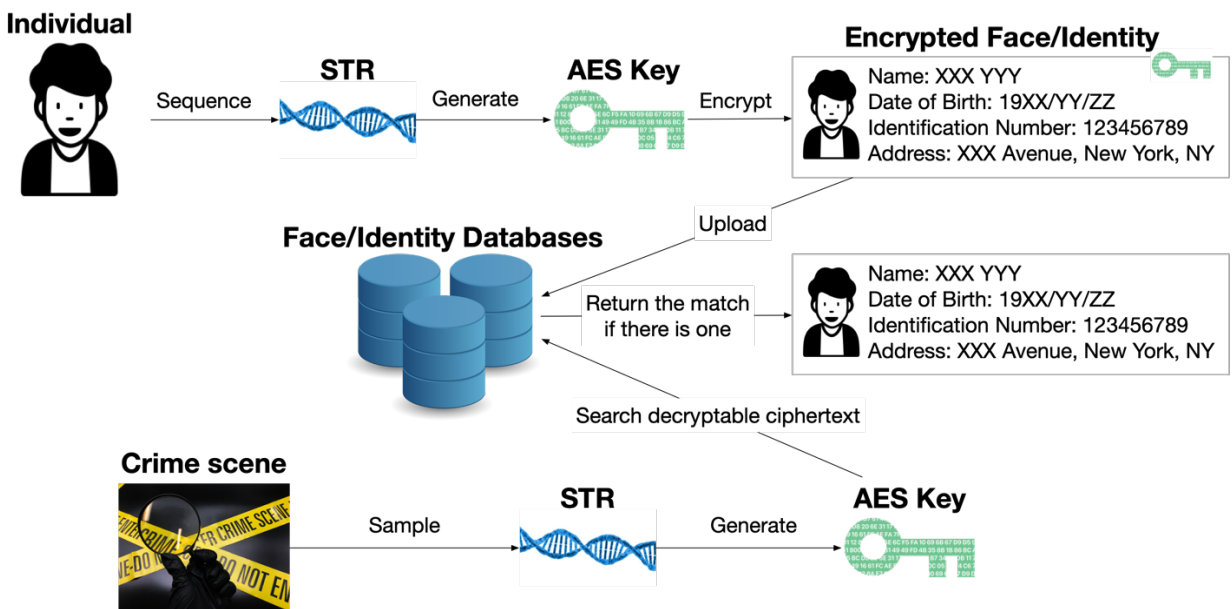

**Figure 1.** Overview of our DNA-based forensic facial recognition

**Toy example of genomic encryption of facial images**
As a toy example (Fig. 2), we used an STR derived key from a recent study investigating the use of STRs in encrypting digital information encoded in synthetic DNA [7]. Note that this previous study is distinct from the framework proposed here as it relies on synthetic DNA that has to be manufactured and mixed with the human DNA [7].

We use the STR key of Individual 1 from Grass et al 2020 [7] to encrypt a sample image (Individual 1, Fig. 2) using the AES algorithm. Then, we decrypt the ciphertext with the same key which yields the original image (i.e., top of Fig. 2). When we use a different key that is randomly chosen to decrypt it, the decryption returns an error message saying "the final block is not properly padded." Note that, even if the final block were properly padded by coincidence, the result would be an incorrect preamble. Thus, STR derived keys can be used to jointly encrypt the STR profile along with biometrics (here a facial image). The resulting ciphertext can then be decrypted back into the biometrics using only the STR derived key.

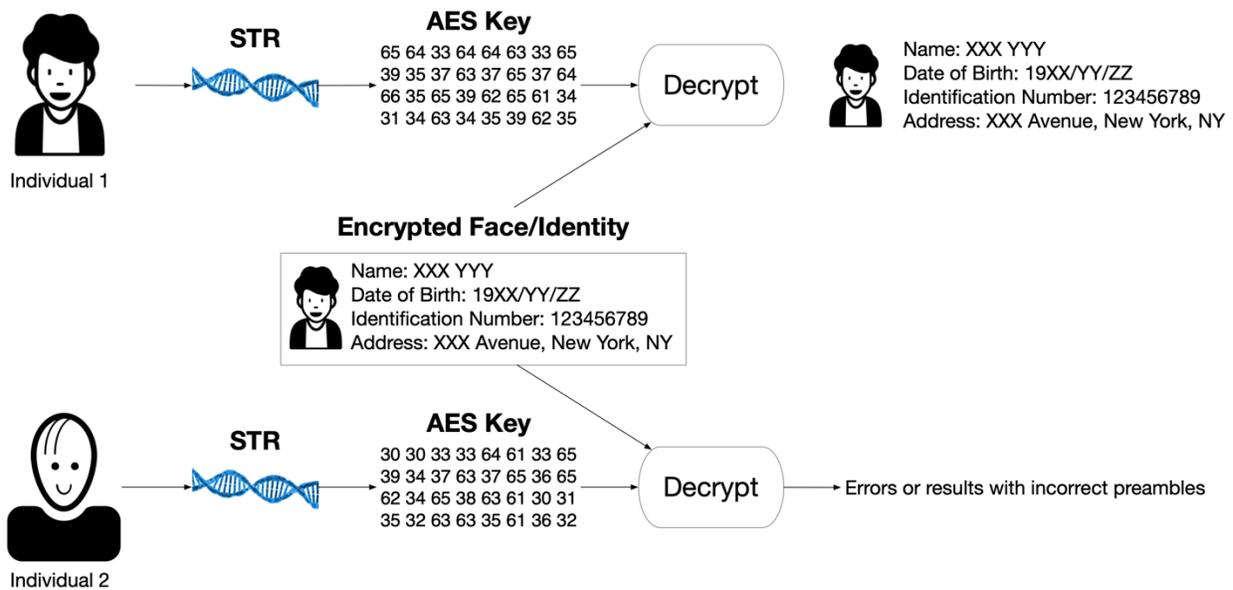

**Figure 2**. A toy example of a match and a non-match

**Security/privacy guarantees**
STRs among different individuals are highly diverse unless they are a direct family (i.e., siblings) [8]. Therefore, law enforcement authorities cannot get an individual's key unless they have gained DNA samples of the individuals. Even a cousin shares only 4 bits in the keys generated from STRs, which reduces the security level of the encryption to 76 bits, which is still secure and would require 10,000 years of computational time at a cost of US $1 billion to break by brute force on a current p3.16xlarge system [7]. Siblings share up to 28 bits in the keys, therefore a sibling can easily compute the key of his/her sibling with brute-force search. However, this only means a sibling can get the face and the identity information of his/her siblings from the database, and the security/privacy impacts from this are limited. This nature of STRs limits the law enforcement officers' access to the records with known DNA/STR information only, meaning that only the records of individuals whose DNA samples are available (e.g., collected from the crime scenes) will be decrypted and accessed. Although it is possible to collect DNA samples from the people with physical contacts (e.g., by collecting saliva or hair of them), STR

sequencing can be strictly controlled and be allowed for authorized purposes only (e.g., allowed for the samples collected from the crime scenes only). Therefore, the records of individuals whose STR do not match any entry in the database will remain secret. To add an additional security guarantee, the STRs -derived keys must be destroyed after their valid usage which would extremely limit the timeframe for a malicious attack. Furthermore, in this system even in the worst-case scenario where the whole database is hacked including at the instance a positive match is found, the data of all individuals in the database remain in ciphertext form hence maintain privacy. Matches are also returned only when a perfect match is found using NIST high stringency conditions [9].

**Accuracy guarantees**

Our approach ties one's STRs to their identity record in the database through the encryption key generated from the STRs. Individuals have highly diverse STRs which provides sufficient entropy for 80-bit keys [7], and only DNA samples from the same person will result in identical STRs. Therefore, an encrypted record can be decrypted correctly if and only if the correct key is used against the ciphertext. Due to the preambles added to the face/identity records, one will see the predefined preamble along with the correct records if and only if the correct key is used. Thus, if the matching is done by searching for decryptable ciphertexts, a match (i.e., a ciphertext is correctly decrypted) is always a correct match and a mismatch (i.e., a ciphertext cannot be correctly decrypted) is always an incorrect match. In other words, the accuracy will be 100% and there will be no false positives or false negatives.

**Efficiency**

According to a benchmark [10], throughput for decrypting a ciphertext with AES-256 can be as high as 1791.73MB/s for a modest server CPU (Intel Xeon E3-1220 V2 with 4 cores at 3.1GHz). Considering that the ciphertexts of face/identity records would have a moderate size (less than 1-5MB), we conjecture the throughput of our genomic encryption framework would be approximately 300-1500 records per second even for a modest server CPU. By leveraging distributed computing with latest server CPUs with tens of cores, one can trivially increase the throughput by two orders of magnitude. For example, if we have 20 server nodes each with Intel Xeon Gold 5320H with 20 cores at 4.2GHz, we will have 400 cores in total. Considering that the search of decryptable ciphertext is perfectly parallelizable, we can expect to increase the throughput by at least 100 times (i.e., 30,000-150,000 records per second). At this throughput, it would take 0.6-3.03 hours to search through 328.2 million people, i.e., the entire population of the U.S and only 1.6-8.3 minutes to search through nearly 15 million offender profiles in CODIS as of February 2021[11].

In conclusion, the proposed framework could be valuable in privacy-preserving human identification using STRs by enabling joint encryption of DNA and other biometric information without having to trust the person running the database. In addition, the use of fully encrypted DNA databases could be robust to malicious attacks as hackers can only access ciphertext at worst. The framework can also be leveraged to safeguard non-biometric data in applications outside of forensics including paternity testing, digital encryption of medical records or data in

DNA biobanks. It could also be leveraged in secure authentication of organs or personalized cell therapies which must be matched to an individual.

**Acknowledgements**
This work was partially supported by the Office of the Director of National Intelligence (ODNI), Intelligence Advanced Research Projects Activity (IARPA) via contract #2020-20082700002 to TJ. Any opinions, findings and conclusions or recommendations expressed in this material are those of the authors and do not necessarily reflect those of the sponsor.

**Conflict of Interest:** The authors are exploring intellectual property in areas related to genomic encryption and privacy-preserving computing on genetic data.